\documentclass{iaus}
\usepackage{natbib}
\usepackage{graphicx}
\def\JApA{JapA} % Journal of Astrophysics and Astronomy

\def\apj{ApJ}%
\def\aap{A\&A}%
\def\mnras{MNRAS}%
\def\acenA{$\alpha$~Cen~A\,}
\newcommand{\eqn} [1] {
\begin{equation}#1
\end{equation}}
\newcommand{\eqna} [1] {
\begin{eqnarray}#1
\end{eqnarray}}

\title[Solar-like oscillation amplitudes and line-widths] %% give here short title %%
{Solar-like oscillation amplitudes and line-widths as a probe for turbulent convection
  in stars}

\author[R. Samadi, K. Belkacem, M.-J. Goupil, F. Kupka and
  M.-A. Dupret]   %% give here short author list %%
{R. Samadi$^1$, K. Belkacem$^1$, M.-J. Goupil$^1$, F. Kupka$^2$ \break
   \and M.-A. Dupret$^1$}

\affiliation{$^1$ Observatoire de Paris, LESIA, Meudon, France; 
  email: reza.samadi@obspm.fr \\[\affilskip]
$^2$ Max-Planck-Institute for Astrophysics, Garching, Germany.
}

\volume{239}  %% insert here IAU Highlights of Astronomy Volume Number
\pubyear{2007}
\pagerange{119--127}                  % or 1--3
\date{?? and in revised form ??}      
\jname{Convection in Astrophysics}
\editors{F. Kupka, I.W. Roxburgh \& K.L. Chan, eds.}

\begin{document}

\maketitle

\begin{abstract}
Excitation of solar-like oscillations is attributed to turbulent
convection and takes place at the upper-most part of the outer convective
zones. Amplitudes of these oscillations depend on the
efficiency of the excitation processes as well as on the properties of turbulent
convection. 
We present past and recent improvements on the modeling of those processes. 
We show how the mode amplitudes and mode line-widths  can bring information
about the turbulence in the specific cases of the Sun and \acenA. 
\end{abstract}

\firstsection % if your document starts with a section,
\section{Introduction}

Solar-like oscillations have now been detected in a dozen main-sequence
stars as well as in some red giant stars \citep[see the recent review by][]{Bedding06}.  Excitation of such
oscillations is ensured by turbulent convection at the upper-most part
of the convective zones.  From the measurement of the mode amplitude
and line-width, it is possible to infer the power supplied to the mode
by turbulent convection.  As highlighted in the
present proceedings, this in turn permits to probe the turbulent
convection in stars. 

Such measurements have been  available for the Sun for several
years and give the possibility to test the
various proposed theoretical models of mode excitation \citep[e.g.][]{GK77,Balmforth92c,Samadi00I,Chaplin05}.
We consider here the model of  \citet[][SG01 hereafter]{Samadi00I} with the improvements
proposed by \citet[][B06b hereafter]{Kevin06b}

The theoretical model of SG01 requires a prescription for the
time-correlation between the turbulent elements. On the basis of a 3D
simulation of the upper part of the solar convective zone,
\citet[][SNS03 hereafter]{Samadi02II} have shown that the eddy
time-correlation, at large scale, is better fitted by a Lorentzian function (LF
hereafter) than a Gaussian function (GF hereafter).
Furthermore, excitation rates of solar p\ modes computed  on the basis of the
model of SG01 with a LF better agree with the excitation rates inferred from
the helioseismic observations by \citet{Chaplin98} than when using a GF. The
open question is whether or not this result remains valid for other stars. 

The theoretical model of stochastic excitation requires a
prescription for the fourth order moments (FOM hereafter)  involving
the entropy fluctuations and the turbulent velocity. 
SG01 assume the quasi-normal approximation (QNA)
which consists in splitting the FOM into the product of two
second-order moments (SOM hereafter). In the solar convective zone,
this simple closure 
model is however significantly biased as verified by the recent
studies performed by \citet[][B06a hereafter]{Kevin06a} and
\citet[][and this volume]{Kupka06}.
Furthermore, as shown in B06b, the skew introduced by the QNA
leads to an under-estimation of the solar p mode excitation rates.  
 
Closure models more sophisticated than the QNA can be used. Among
those, the so-called two-scale mass flux model (TFM hereafter) improved by
\citet{Gryanik02} takes  
the asymmetries in the medium  into account but is only applicable for quasi-laminar
flows. B06a  have  generalized
\citet{Gryanik02}'s  approach by taking  the turbulent properties of
the medium into  account. In order to apply
this model in the solar case, they have introduced the plumes dynamics
following \citet{Rieutord95}.	
Furthermore, as shown in B06b, the calculations based on this new closure
model increases the contribution of the Reynolds stress contribution
to the excitation rates of the solar modes. 
When the additional contribution due to the entropy fluctuations is
included, the new theoretical calculations fit rather well  the
maximum in the solar mode excitation rates derived  by \citet{Baudin05}. 
All the results by B06b are summarized and discussed in the present
proceedings (see also Belkacem et~al., this volume).  
 
Apart from the Sun, mode amplitudes \emph{and} mode line-widths
have been derived for few stars. 
Among those stars, \acenA  provides the best available data.
\citet{Bedding04} have derived the mode amplitudes and the
mode line-widths for \acenA.
These mode line-widths have been recently updated by  \citet{Kjeldsen05}.
More recently, \citet{Fletcher06} have derived the mode
line-widths from the WIRE observations of  \acenA \citep{Schou01}.
From those new measurements we have updated the work performed in
\citet{Samadi04b} by deriving new constraints on the rate at
which energy is supplied by unit time to the  p~modes of \acenA.
These constraints are compared here with new theoretical
calculations performed in the manner of B06b.

This report is organized as follows: in \S~\ref{model} we briefly recall
the model of SG01 with the modifications proposed by B06b.
We also present the different ways the eddy-time correlation is described and
the different closure models investigated by B06a.
Applications of this improved  excitation model are shown in \S~\ref{applications} to the case of the Sun
(\S~\ref{sun}) and of \acenA (\S~\ref{acena}).
Finally \S~\ref{conclusion} and \S~\ref{discussion} are dedicated to
conclusion and discussion respectively.  
%and of different
% stars for which solar-like oscillations  have been up to now
% detected (\S~\ref{stars}).

\section{Modeling the stochastic excitation}
\label{model}

The model of stochastic excitation we consider  basically is that of SG01 with
the improvements proposed by B06b: two terms are expected to drive
stochastically the p~modes: one term corresponds to the Reynolds
stress, the second one is an entropy source term that corresponds to
the advection of entropy fluctuations by turbulent motions. 
The energy supply per unit
time into each mode is given by SG01: 
\eqn{  P  =  \frac{1}{8 \, I} \left ( C_R^2 +  C_S^2 \right ) 
\label{eqn:power} }
where $C_R^2$ and $C_S^2$ are the turbulent Reynolds stress and
entropy contributions respectively. Their expressions are for radial
modes: 
\eqna{
C_R^2  =   \int_{M}  d m \, \rho_0 f_r   \int_{-\infty}^{+\infty} d\tau \, e^{-i\omega_0 \tau} \, \int d^3r \, \left <
 w_1^2  w_2^2
\right >   [\vec r, \tau]
\label{C2R_rad} \\
C_S^2  =    \int_{M}  d m  \, \alpha_s ^2 \, { g_r  \over \rho_0} \int_{-\infty}^{+\infty} d\tau \, e^{-i\omega_0 \tau}
\int d^3r \, \left <
 \left ( w s_t     \right)_1
\left ( w s_t  \right )_2  \,
 \right >  [\vec r, \tau]
\label{C2S_rad}
}
where $\rho_0$ is the mean density, $w$  the vertical component of the velocity, $s_t$ the entropy
fluctuation due to turbulence, $\alpha_s \equiv \partial (p /
\partial s)_\rho$,  $m$ the mass enclosed in a sphere with radius $r$, $I$ the mode inertia, $f_r(\xi_r,m) \equiv \left
({\partial \xi_r \over \partial r}\right )^2$, $\xi_r$ the
radial component of the displacement eigenfunction, $g_r$ is a function
that involves the first and second derivatives of $\xi_r$ \citep[see Eq.\ (9)
of][]{Samadi02I}, finally $\tau$ and $r$ are variables related to
the time and space  correlation lengths respectively.

Quantities labeled with  1 and 2 are taken at the spatial and temporal
positions $[ \vec x_0-\frac{\vec r}{2}, - \frac{\tau} {2}]$
and $ [\vec x_0+\frac{\vec r}{2}, \frac{\tau} {2}]$ respectively:
they correspond to {\it two-point} correlation products. 
The  fourth-order two-point correlation product involving the velocity,$< w_1^2
w_2^2>$, is related to the fourth-order  {\it one-point} correlation
product, $< w_1^4> $, as summarized in \S~\ref{2pt}. 
We point out that the fourth-order one-point correlation
product, $< w_1^4>$, corresponds to a FOM in the usual sense of Reynolds
averages. In turn, the one-point correlation product, {\it i.e.}
the FOM, is modeled as described in \S~\ref{closure}.

\subsection{Closure models}
\label{closure}

\begin{figure}
\parbox{8.25cm}{
 \includegraphics[width=8cm]{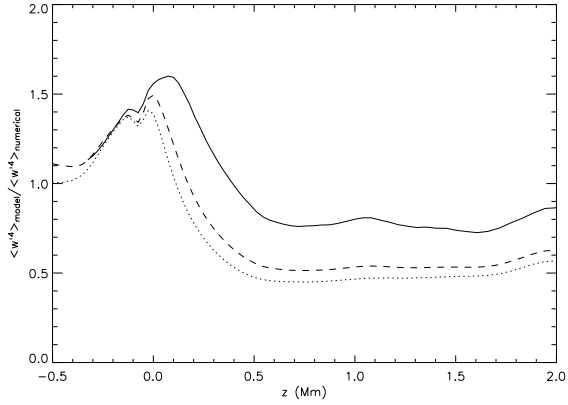} 
}
\parbox{5cm}{
 \caption{Fourth-order moment (FOM) of the velocity, $<w^4>$, as a
    function of depth $z$, normalized to the FOM derived from the 3D
    simulation. In solid lines the FOM
    calculated according to the CMP model, {\it i.e.} with the help of
    Eq.~\ref{m4_GH} and  with the
    skewness, $S_w$, given by Eq.~\ref{Sw_cmp}, the dashed line is 
    computed according to the \citet{Gryanik02}'s TFM given by 
    Eq.~\ref{m4_GH} with $S_w$ given by with
    Eq.~\ref{Sw_GH} and finally the dotted line is the QNA, Eq.~\ref{m4_qna}.
}
}
\label{fig:fom}
\end{figure}

The QNA \citep[see][Chap VII-2]{Lesieur97} allows us to calculate rather
easily the FOM of $w$ in terms of a product of the SOM, that is: 
\eqna{
  \langle w^4 \rangle  = 3 \, \langle w^2 \rangle^2   \; .
\label{m4_qna}
}
The QNA is strictly valid for normally distributed
fluctuating quantities with zero mean. However, the upper-most part of the convection zone being a turbulent
convective system composed of essentially two flows, the probability distribution
function of the fluctuations of the vertical velocity and temperature
do not follow a gaussian law \citep[see e.g.][B06a]{Lesieur97}. 
Hence the use of the QNA is not really valid.
This was verified by B06a and \citet{Kupka06} with the help of  3D simulations of the
outer layer of  the Sun:  
Figure~\ref{fig:fom} shows  the ratio between the FOM $<w^4>$ derived
from a 3D simulation as detailled in B06a and the FOM derived from
different closure models. 
As shown in Figure~\ref{fig:fom}, the use of the QNA
under estimates, in the quasi-adiabatic region,  by
$\approx$~50~\% the FOM of the velocity derived from 
the 3D simulation.

A more sophisticated  closure model, the  mass flux model
\citep[][MFM hereafter]{Abdella97},
takes the effects of {\it updrafts} and {\it downdrafts}
on the correlation products into account. Indeed, the presence of two flows
introduces a skew when averaging the fluctuating quantities, since averages
of fluctuating quantites over each individual flows differ from averages over
the total flow. The basic idea of the MFM is to split any averaged turbulent
quantity $\phi$ into two parts, one associated with the updrafts and the other
with the downdrafts.
\citet{Gryanik02} have proposed their so-called two-scale mass-flux
model (TFM hereafter) which extends the MFM  by accounting for the fact that
hot drafts (or cold ones, resp.) do not necessarily coincide with the updrafts
(downdrafts, resp.). In order to close the third order moments and the FOMs,
\citet{Gryanik02} adopt the simplifying approximation that
$<\phi^n> = <\phi>^n$ where $<.>$ denotes ensemble spatial (in the
horizontal plane) and time averages. Hence, this
approximation consists in neglecting the turbulence within the flow.
Finally, \citet{Gryanik02}  derive  analytical
expressions for the third-order moments and FOMs involving the temperature
and the velocity. For $<w^4>$, \citet{Gryanik02} obtained the following
expression:
\eqna{
\label{m4_TFM}
 <w^4>          &=&  (1+S^2_{w}) ~<w^2>^2 
}
with  the skewness, $S_w$, given by:
\begin{equation}
\label{Sw_GH}
  S_{w} \equiv {<w^3> \over <w>^{3/2} } = \frac{1-2a}{\sqrt{a(1-a)}} \; 
\end{equation}
where $a$ is the mean fractional area occupied by the updrafts in the
horizontal plane. In the QNA limit, {\it i.e.} when the random quantities
are distributed according to a normal distribution with zero mean, we have
necessarily $S_w=0$. Hence, in the QNA limit, Eq.~\ref{m4_TFM} does not match
Eq.~\ref{m4_qna}. \citet{Gryanik02} proposed to modify Eq.~\ref{m4_TFM} as
follows (see \citealt{Kupka06} for a discussion of this step):
\eqna{
\label{m4_GH}
 <w^4>          &=&  3 \, (1+{1 \over3} \, S^2_{w}) ~<w^2>^2.
}
Figure~\ref{fig:fom} shows that the FOM  based on
Eq.~\ref{m4_GH} with  $S_w$ given by Eq.~\ref{Sw_GH}, results in a negligible
improvement with respect to the QNA. 
However, when  $S_w$ is derived directly from the 3D simulation and
plugged into Eq.\ \ref{m4_GH}, Eq.~\ref{m4_GH} is a very good evaluation
of the FOM derived from a 3D simuation of the outer layer of the Sun as
verfied by B06a and \citet[][]{Kupka06}.

B06a have generalized the TFM  by taking both the skew
introduced by the presence of two flows 
\emph{and} the effects of turbulence inside each flow into account.
Accordingly, they have derived   a more accurate expression for $S_w$,
that is: 
\begin{equation}
\label{Sw_cmp}
  S_{w}=  \frac{1}{<w^2>^{3/2}}~ a~  \Big((1-a)(1- 5a)\, \delta w^2-  3 <w^2> \Big) \delta w
\end{equation}
where $\delta w$ is the difference between the average velocity of the
up- and the downdrafts and $<w^2>$ is the SOM. Both
$a$ and  $<w^2>$ are supposed to be known. Finally, in order to close
the system, $\delta w$ needs to be modeled. This is undertaken in B06a
by using the plume model of \citet{Rieutord95}. More details can
be found in B06a or in \citet[][]{Kevin06d}.

\subsection{Two-point and Eddy-time correlations}
\label{2pt}

As seen in Eq.~\ref{C2R_rad} and Eq.~\ref{C2S_rad}, the model of
stochastic excitation relies on a prescription for the two point
correlation products involving the velocity (Eq.~\ref{C2R_rad} and
Eq.~\ref{C2S_rad}) and the entropy (Eq.~\ref{C2S_rad}).
B06b proposed to relate the two-point correlation products involving the
velocity to the one-point correlation with the help of QNA as follows:
\eqn{
<w^2_1 w^2_2> = {  <w^4> \over <w^4>_{\rm QNA} } \, <w^2_1 w^2_2>_{\rm QNA}
\label{2ptcorel}
}
where $ <w^4> /  <w^4>_{\rm QNA}$ is the ratio betwen the FOM of
 $w$  and the one given by the QNA (Eq.~\ref{m4_qna}), 
$<w^2_1 w^2_2>_{\rm QNA}$ is the two-point correlation product given
according to the QNA \citep[see][Chap VII-2]{Lesieur97}, that is:
\eqn{
  \langle w_1^2 w_2^2 \rangle_{\rm QNA}   = 2 \, \langle w_1 w_2  \rangle^2 +  \langle w_1^2  \rangle \langle w_2^2 \rangle  \; .
\label{2pt:qna}
} 
Note that the second term in Eq.~\ref{2pt:qna} does not contribute to
the excitation. The FOM in Eq.~\ref{2ptcorel} is here computed
according to the closure models presented in \S~\ref{closure}.
When the CMP is adopted, from Eqs.~\ref{2ptcorel}, Eq.~\ref{m4_GH} and
Eq.~\ref{m4_qna} one  derives the following
expression for $<w^2_1 w^2_2>$ (Eq.~7 of B06b):
\eqn{
<w^2_1 w^2_2> = (1+\frac{1}{3} S_{w}^2) <w^2_1 w^2_2>_{\rm QNA} 
\label{2pt:cmp}
}
where $<w^2_1 w^2_2>_{\rm QNA}$ is given by Eq.~\ref{2pt:qna}.

SG01 introduce  $\phi_{ij}( \vec k,\omega)$ the time and space Fourier
transform of  $<w_1 w_2>_{\rm QNA}$ (the first term in Eq.~\ref{2pt:qna}).
SG01 modeled  $\phi_{ij}( \vec k,\omega)$ as follows:
\eqna{
\phi_{ij}( \vec k,\omega)  =  \frac { E( k,\omega) } { 4 \pi k^2}
\left( \delta_{ij}- \frac {k_i k_j} {k^2}  \right) & \; {\rm with} \; & E( k,\omega) =E(  k) \, \chi_k(\omega)
\label{eqn:phi_ij}
} 
where $ E( k,\omega) $ is the turbulent kinetic energy spectrum, $\delta_{ij}$ 
the Kroenecker symbol, $E(k)$ the mean kinetic energy spectrum and
$\chi_k(\omega)$ the eddy-time correlation function.
Several eddy-time correlation functions have been investigated by
SNS03. Among those, we compare here the Gaussian function (GF
hereafter) and the
Lorentzian function (LF hereafter).
Finally from Eqs.~\ref{2ptcorel}-\ref{eqn:phi_ij}, B06b derive the CMP
version of Eq.~\ref{C2R_rad} (that is Eq.~10 of B06b). Note that as in
SG01 and B06b, the entropy source term (Eq. \ref{C2S_rad}) is still modelled on
the basis of the QNA.

\section{Results}
\label{applications}

\subsection{The Sun}
\label{sun}

The calculations of the excitation of the solar modes is performed as
detailed in B06b: the radial eigenfunctions,
$\xi_{\rm r}$, eigenfrequencies, $\omega_0$, and the mode inertia, $I$,
are those used by  \citet{Samadi02I} and originate from 
an 1D solar model built  according to Gough's (1977)\nocite{Gough77} non-local
formulation of  the mixing-length theory. 
The spatial and time averaged quantities  are  obtained  from a 3D
simulation of the solar surface computed with \cite{Stein98}'s code
and with  a grid of $125 \times 125 \times  82$.
The total kinetic energy, $E(k)$, in Eq.~\ref{eqn:phi_ij} and its depth
dependence  are obtained directly from the 3D simulation. Finally,
quantities related to the convection (the SOM of $w$ and  $s_t$, and
$a$) and thermodynamic quantities (the mean pressure $p_0$, $\rho_0$ and
$\alpha_s$) are also obtained from the 3D simulation.
The results of the calculations of the mode excitation rates (Eq.\ref{eqn:power})
are shown for the Sun in Figure~\ref{fig:sun}, top panel.
 
\subsection{$\alpha$~Cen~A}
\label{acena}

\begin{figure}
\parbox{8.25cm}{
 \includegraphics[width=8cm]{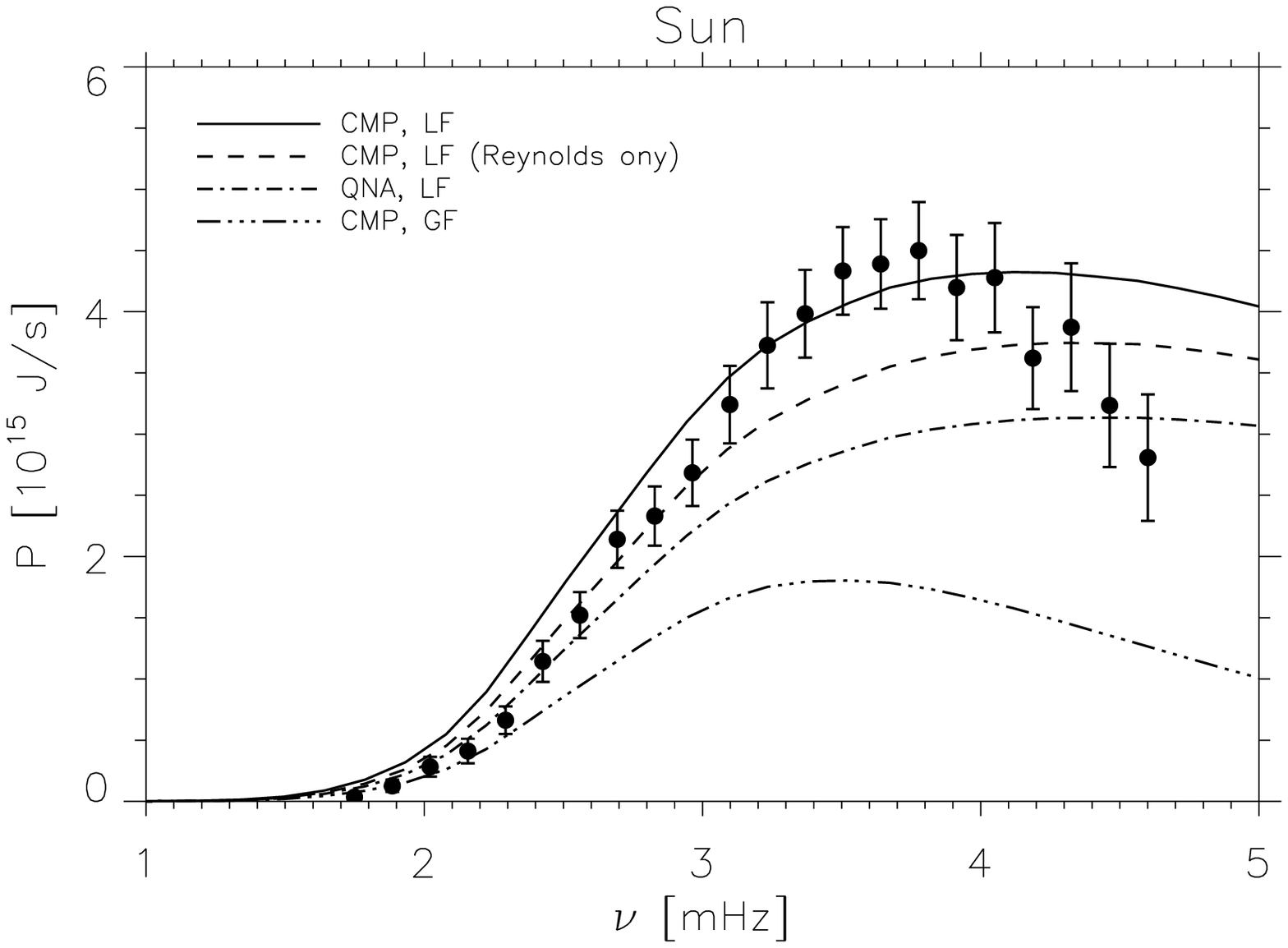}\\
 \includegraphics[width=8cm]{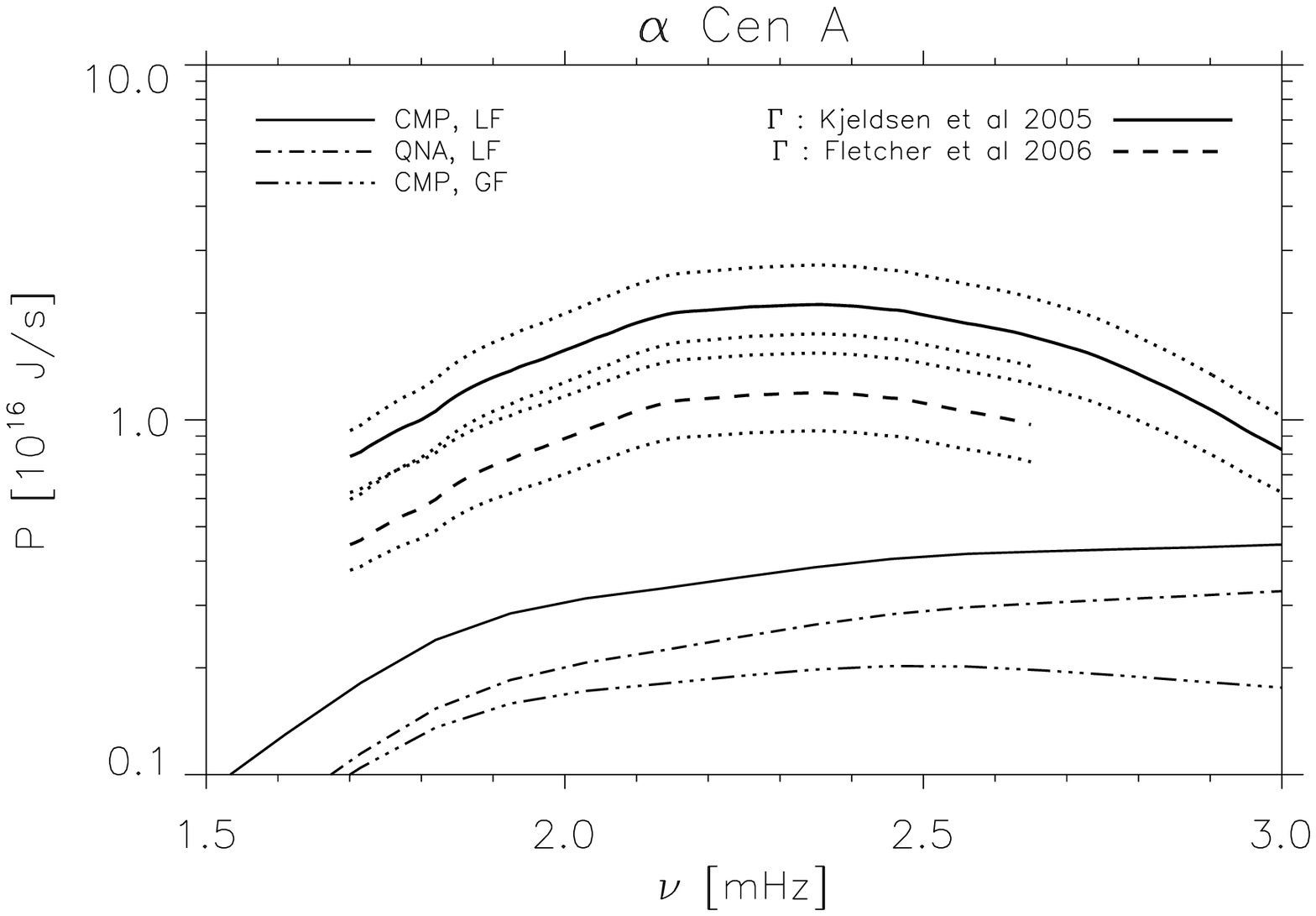}

}
  \parbox{5cm}{
  \caption{{\it Top:} Rates at which energy is injected into the solar modes.
    The filled circles correspond to the helioseismic constraints obtained 
    by \citet{Baudin05}. The lines correspond to different theoretical
    calculations: the solid line uses the Lorentzian function (LF) and the CMP,
    the dashed line is as the solid line with only the
    contribution of the Reynolds stress, 
    the dot-dashed line uses the LF and the QNA closure model, the
    triple-dot-dashed line uses the Gaussian function (GF) and the CMP.
    {\it Bottom:} Same as top for \acenA. The thick lines correspond to
    the constraints obtained from the observed spectrum derived by
    \citet{Kjeldsen05} and the averaged mode line-widths derived by
    \citet[][the thick solid line]{Kjeldsen05}  and by 
    \citet[][the thick dashed line]{Fletcher06}.  The thick dotted
    lines correspond to extreme values derived from the estimated
    errors on the amplitude and line-width measurements.
    The thin lines have the same meaning as in the top panel.
}
}
\label{fig:sun}
\label{fig:acena}
\end{figure}

Calculations of mode excitation rates are performed for \acenA in a
similar way as for the Sun: the simulation has a grid of
$125 \times 125 \times  82$, an effective temperature of $T_{\rm eff}=5805
\pm 20$~K and a gravity of $\log g = 4.305$ in agreement with the constraints
derived from interferometry \citep[see][]{Miglio05}. We compute a 1D model
with the CESAM code assuming standard physics and the same $T_{\rm eff}$ and
$\log g$ as for the simulation. Finally, $\xi_{\rm r}$, $\omega_0$ and $I$
are computed with an adiabatic code from the 1D model.
The results of the calculations of the mode excitation rates 
are shown for \acenA in Figure~\ref{fig:sun}, bottom panel.
The seismic constraints are derived as in
\citet{Samadi04b} except that, after adding the apparent amplitudes for
$\ell$ to those for $\ell+2$ (with $\ell=0$ and 1), we divide the
summed amplitudes by the sum of the visibility factors of all the
summed modes including the multiplets. Furthermore,  the mode masses
(${\cal M} = I/ \xi_{\rm r}^2$) are evaluated as in
\citet{Samadi04b} at the photosphere.

%Teff = 5804.66469 K
%logg = 4.30489893
% M =1.105 M_0
%alpha=1.3697355, age=4.94491897 Gy

\section{Conclusion}
\label{conclusion}

Comparison between  recent helioseismic constraints obtained by
\citet{Baudin05} and the theoretical calculations based on the
quasi-normal approximation (QNA) confirms the result of SNS03 that
the Lorentzian function (LF) results in a better agreement with the
observations than the Gaussian function (GF). 

As for the Sun, we find that for \acenA the  theoretical
calculations based on a Lorentzian eddy-time correlation function   is
closer to the observations than those based on  a Gaussian one.

In the case of the Sun,  B06b have shown that the
QNA results in a significant underestimation
of the contribution of Reynolds stress to the mode excitation. 
On the other hand, the so-called closure model with plumes (CMP)
proposed by B06a increases significantly this 
contribution. The theoretical calculations limited to the
Reynolds stress only underestimates by $\sim$~15 \% the maximum in
the excitation rates $P$  derived by \citet{Baudin05}. 
On the other hand, when the entropy  contribution is added, the
theoretical calculations based on this
improved closure model fit the helioseismic data. 

In the case of \acenA, the theoretical calculations underestimate
the observations by a factor ten. We note also that the accuracy at which 
mode amplitudes and line-widths are measured is more or less of the
same order as the difference between theoretical calculations using
the QNA and those using the CMP. This emphasises the need for more
accurate seismic data.

Whatever the assumptions about the eddy-time correlation or the
closure model, the theoretical calculations are found to be closer to the
constraints derived using the line-widths of \citet{Fletcher06} than to
those obtained for \citet{Kjeldsen05}'s.

The space based mission CoRoT (see the recent review by \citealt{Michel06})
is the only asteroseismology mission that in the very near future will 
enable us to derive, for a large set of solar-like oscillating stars
with different effective temperature and gravity, the rates at which
energy is supplied to the modes by turbulent convection. 
The quality of the data is expected to be significantly higher than
current observations.
Hence, these data will thus  very likely permit to discriminate
between the best description for the eddy-time correlation function
and closure models.

\section{Discussion}
\label{discussion}

For the Sun, the largest discrepancies between theoretical
calculations and the observations are seen at high frequency ($\nu
\gtrsim $ 4 mHz) and at low  frequency ($\nu \lesssim$ 3 mHz).

Part of the remaining discrepancies with the helioseismic data can be
attributed to the closure models considered here which are not able to reproduce
correctly the FOM from the 3D simulation in the super-adiabatic region. Indeed,
as seen in Figure~\ref{fig:fom}, all the closure models considered here
overestimate the FOM. B06b have shown that the main effect on the excitation
rates is at rather high frequency and results in an increase of the
order of $\sim$\ 15~\% when the CMP is adopted.  

In B06b, the one-point correlation of the velocity was generalized  to
a two-point correlation product using the QNA (see \S~\ref{2pt}). It
was shown in B06b that, at small scale length, this approximation
reproduces satisfactorily the constraints derived from the 3D
simulation. At large scale length, however, this approximation
overestimates those constraints. The effect on the excitation rates of this
overestimation at large scale  remains to be evaluated. 

Finally, we must point out that the calculation of the mode excitation
rates requires the calculation of the mode eigenfunctions. The latter
are obtained on the basis of global 1D stellar models. In such stellar
models, the structure of the outer layer is very poorly
modeled. Therefore, this poor description must have some (currently
unknown) consequences on the inferred properties of the  mode
eigenfunctions in particular on the high frequency modes which are
almost confined near the photosphere. In turn, as most of the excitation
occurs near the photosphere, this poor description is likely to
introduce some (unknown) biases on the predicted excitation rates in
particular at high frequency. 

The discrepancies at low frequency can probably be attributed to
the adopted model for the eddy time-correlation which is probably too simple.
Indeed, the departures from a Gaussian
behavior were attributed in SNS03 to the plumes, which are more
turbulent than the granules. Near the photosphere, the excitation of
the p\ modes is almost solely due to the plumes. However, one Mm deeper, the
excitation is no longer dominated by the plumes.  As the granules are
less turbulent than the plumes, their time-correlations
follow more likely a GF distribution than a LF. Then one Mm deeper the
eddy-time correlation is expected to lie between the LF and the GF. This
however remains to be checked.   

For \acenA, the large discrepancies between the observations and the theoretical
calculations remain to be understood. In addition to what is mentioned above for
the case of the Sun, the ways the mode excitation rates were computed in the
particular case of \acenA may have some (unknown) effects on the computed
excitation rates. Indeed, the adiabatic eigenfunctions were obtained from a
1D model based on the local mixing-length theory. Finally, the 3D
simulation considered here uses solar abundances.
On the observational side, we shall mention that the excitation rates
were derived from the observations using mode masses evaluated
arbitrarily at the photosphere. As the doppler measurements used a lot
of spectral lines, it is not obvious to evaluate the observational
mode mass.

The work performed in SNS03 for the case of the Sun has been extended
by \citet{Samadi05c} to the case of
stars lying on the main sequence for which solar-like 
oscillations are expected. They have found that the maximum in
the mode amplitudes, $V_{\rm max}$, scales as $(L/M)^s$ where
$s$ is the slope of the scaling law, $L$ is 
the luminosity and $M$ is the mass of the star.
The  slope $s$ was found to be significantly
sensitive to the adopted eddy-time correlation function.
As for the Sun and \acenA, theoretical calculations assuming an
LF fits best the maximum oscillation amplitudes
detected for a set of $\sim$~10 stars including main-sequence stars as
well as some red giant stars. 
However, these calculations rely on the QNA. A preliminary work tends to
show that the asymmetry between the updrafts and the downdrafts does not
change significantly between the few  3D simulations of main-sequence
stars investigated in this work and that the CMP remains valid
according to these 3D simulations. Hence, we expect that the difference
between the effect on the excitation rates of the CMP model and that of
the QNA remains constant for intermediate massive stars lying on the main
sequence. This however needs to be confirmed using more 3D simulations
and extended to other domains of the HR diagram.

\begin{acknowledgments}
FK is grateful to V.M.~Gryanik and J.~Hartmann for discussions on their model and their observational data. 
We thank A. Nordlund and R.~F. Stein for making their code available
to us.  Their code was made at the National Center for Supercomputer
Applications and Michigan State University and supported by grants
from NASA and NSF.
\end{acknowledgments}

%\bibliography{../../../biblio}

\bibliographystyle{aa}

\begin{discussion}

\discuss{V.M. Canuto}{How do you determine the filling factor~?}

\discuss{R. Samadi}{This quantity is obtained directly from the 3D
simulation.}

\discuss{Tim Bedding}{Comment: It is very nice that you can compare your models
with our observations. But please do not compare with amplitudes of individual
peaks. There is a systematic effect because those peaks are selected
because they happen to be highest during the observations. (This is true
even if you average over many peaks). Instead, please use the amplitudes
we estimated by smoothing and then noise-correcting the power spectrum.
For $\alpha$~Cen A and B, this is shown in Kjeldsen et\ al. (2005).}

\discuss{R. Samadi}{Yes, I agree with you, the comparison should be done
using the type of amplitude spectrum you mention. (Thanks to
T. Bedding this is now done in the present proceedings).}

\discuss{R.F. Stein}{First, we have compared the non-adiabatic eigenfunctions
from our simulations with the eigenfunctions of J. Christensen-Dalsgaard
and they agree very closely, so I don't think that using non-adiabatic
eigenfunctions is important. Second, you add the absolute values
squared of the turbulent pressure and entropy fluctuation
contributions. In fact, they do not always add but sometimes cancel
each other. Hence, adding the contributions linearly would be more
accurate.}

\discuss{R. Samadi}{We have compared the non-adiabatic eigenfunctions
computed using the time-dependent formalism of Gabriel for convection (see
Grigahcene et al.\ 2005) with adiabatic eigenfunctions and found
important differences. Hence, different treatments of convection affect
differently the eigenfunctions. Further work is thus needed on that issue.
Concerning your remark about the turbulent pressure and entropy fluctuation,
within the theoretical framework of our formalism, the crossing
term between the Reynolds stress source term and the entropy source
term vanishes on average. Hence, further theoretical work is required in
order to correctly model this crossing term.}

\discuss{J. Christensen-Dalsgaard}{Comment: The computations of damping rates
(needed to determine the amplitude) is probably even more uncertain than the
computation of the energy input. For the red giant $\xi$~Hya there is
a striking discrepancy between observed and computed damping
rates. Better understanding, and better observations, are needed.}

\end{discussion}

\end{document}